\documentclass[twocolumn]{aastex631}

\usepackage{float}
\usepackage{multirow}
\usepackage{threeparttable}
\usepackage{tabularx}
\usepackage{amsmath}
\usepackage{hyperref}

\shorttitle{PKS 1510-089: II. Jet-BLR Connection and Black Hole Mass Estimation}
\shortauthors{Amador-Portes et al.}
\graphicspath{{./}{figures/}}
\begin{document}

\title{Unveiling the Emission Mechanisms of Blazar PKS 1510-089: II. Jet-BLR Connection and Black Hole Mass Estimation}

\correspondingauthor{Alfredo Amador-Portes}
\email{alfre\_portess97@hotmail.com,aamador@inaoe.com}

\author[0009-0009-6341-0270]{Alfredo Amador-Portes}
\affiliation{Instituto Nacional de Astrofísica, Óptica y Electrónica, Luis Enrique Erro \#1, Tonantzintla Puebla, México, C.P. 72840}

\author[0000-0002-2558-0967]{Vahram Chavushyan}
\affiliation{Instituto Nacional de Astrofísica, Óptica y Electrónica, Luis Enrique Erro \#1, Tonantzintla Puebla, México, C.P. 72840}

\author[0000-0002-5442-818X]{Víctor M. Patiño-Álvarez}
\affiliation{Instituto Nacional de Astrofísica, Óptica y Electrónica, Luis Enrique Erro \#1, Tonantzintla Puebla, México, C.P. 72840}
\affiliation{Max-Planck-Institut für Radioastronomie, Auf dem Hügel 69, D-53121 Bonn, Germany}

\author[0000-0003-4513-7863]{José Ramón-Valdés}
\affiliation{Instituto Nacional de Astrofísica, Óptica y Electrónica, Luis Enrique Erro \#1, Tonantzintla Puebla, México, C.P. 72840}

\begin{abstract}
The flat spectrum radio quasar PKS 1510-089 is one of the most active blazars across the entire electromagnetic spectrum, displaying periods of flaring activity. This study explores its spectral variability over a decade. By employing the non-thermal dominance parameter, we analyze the H$\beta$ and $\lambda5100\text{ \AA}$ continuum light curves, as well as the full width at half maximum of the H$\beta$ emission line, to identify whether the primary source of the continuum emission is the accretion disk or the jet during activity periods. Our results shows an anti-correlation between the full width at half maximum and the luminosity of the H$\beta$ emission line across all datasets. This indicates, that variations in H$\beta$ luminosity consistently reflects the canonical broad-line region, irrespective of whether the primary ionizing source is the accretion disk or the jet. The anti-correlation persisted when comparing the full width at half maximum of H$\beta$ against the luminosity at $\lambda5100\text{ \AA}$ in the disk dominance regime. These findings, along with the observation that flaring events in the $\lambda5100\text{ \AA}$ continuum, attributed to the jet, coincide with flares in the H$\beta$ emission line, suggest that the base of the jet is located within the broad-line region. Based on the 219 spectra within the disk dominance regime, we estimated a mean black hole mass of $M_{BH}=2.85\pm0.37\times10^{8}\: M_{\odot}$.
\end{abstract}


\section{Introduction} \label{sec:intro}

The broad-line region (BLR) is a fundamental component of active galactic nuclei (AGNs). Situated close to the central supermassive black hole (SMBH; \citealp{UrryAndPadovani1995}), the BLR is photo-ionized by ultraviolet (UV) and optical photons from the accretion disk, resulting in the presence of broad emission lines in the UV/optical spectra. With a range of typical full width at half maximum (FWHM) values of $10^{3}-10^{4}$ km s$^{-1}$ \citep{Peterson1993}. Properties of the BLR, such as the FWHM, the luminosity of their emission lines, and the delay between emission line and continuum fluxes, serve as tracers that provide information about the kinematics of the BLR and the mass of the black hole (M$_{BH}$) at the core of the central engine.

Since the BLR is ionized with photons from the central engine, variations in the continuum flux will be followed by the subsequent variations in the emission line fluxes. Therefore, the size and velocity structure of the BLR vary in response to changes in the ionizing continuum flux from the accretion disk, this is known as the ``breathing-BLR'' effect (e.g. \citealp{Peterson1992, Peterson2002, Korista2004, Park2012, Barth2015}). Techniques like reverberation mapping (RM; \citealp{Peterson1993, Kaspi2000}) measure this time delay (or lag) between variations, providing an estimate of the size of the BLR, and when combined with a velocity parameter (FWHM) of the broad emission lines, a virial mass estimation can be derived for the SMBH. Empirical scaling relationships have been developed from objects in which it has been possible to apply RM, allowing M$_{BH}$ estimates using single-epoch spectra. These relationships rely on the luminosity of the AGN continuum and the FWHM of specific broad emission lines (e.g. \citealp{Kong2006, VestergaardAndPeterson2006, Shen2011, Shaw2012}). For example, the H$\beta$ line is often used, where its FWHM and the luminosity at $\lambda5100\text{ \AA}$ (or the H$\beta$ luminosity) are used to estimate the M$_{BH}$ \citep{GreeneAndHo2005}.

Accurately measuring those dynamical tracers (and consequently, estimating the M$_{BH}$) in radio-loud AGNs is significantly more challenging. This is because their optical emission is overwhelmed by the non-thermal radiation from their relativistic jets. Which in the case
of blazars, are aligned close to our line of sight \citep{UrryAndPadovani1995}. Therefore, it is possible that the jet UV/continuum emission can serve as an ionizing source for the BLR clouds, disrupting the scenario of a virialized BLR from which RM and single-epoch methods arise. It is then that additional considerations need to be taken into account to use RM or single-epoch. Previous works have addressed scenarios with BLR clouds ionized by jet emission to explain quasi-simultaneous optical continuum and emission line variability for a sample of blazars \citep{Perez1989} and also for specific sources, e.g. 3C 454.3 \citep{LeonTavares2013, Isler2013, Jorstad2013},  3C 273 \citep{PaltaniAndTurler2003}, CTA 102 \citep{Chavushyan2020}, and Ton 599 \citep{Hallum2022}. 

The source of study is the flat-spectrum radio quasar (FSRQ) PKS 1510-089, (redshift $z=0.361$; \citealp{BurbidgeAndKinman1966}), notable for its high variability across the entire electromagnetic spectrum (e.g. \citealp{Marscher2010, Rani2010, Aleksic2014, Fuhrmann2016, Prince2019, Yuan2023}). The relativistic jet in PKS 1510-089 is nearly aimed directly at us, with an angle of approximately 3 degrees from our line of sight \citep{Homan2002}. Allowing us to observe features within the jet traveling at apparent speeds as high as 20 times the speed of light ($20c$; \citealp{Jorstad2005}). In \citeauthor{AmadorPortes2024b} (\citeyear{AmadorPortes2024b}, hereafter \hyperlink{paperI}{Paper I}), we found that during flaring events, the continuum emission at $\lambda5100\text{ \AA}$ continuum, J-band, and V-band are primarily due to synchrotron emission from the jet. Additionally, we observed an  $\sim80$-day delay between the continuum emission at $\lambda5100\text{ \AA}$ continuum and the H$\beta$ emission line flux. 

The spectral properties of PKS 1510-089 have been explored to a limited extent, due to the high Fe II emission in their spectra, along with the telluric absorptions between the H$\beta$ and [O III]$\lambda\lambda4959,5007\text{ \AA}$ emission lines. \citet{Brotherton1996} found (as part of a sample of 60 blazars) a correlation between the $\lambda5100\text{ \AA}$ continuum luminosity and [O III]$\lambda5007\text{ \AA}$ line width and an anti-correlation between the $\lambda5100\text{ \AA}$ continuum luminosity and the equivalent width (EW)  of H$\beta$ and [O III]$\lambda5007\text{ \AA}$ (i.e. the Baldwin effect). Also, they indicate the presence of a very broad component in the H$\beta$ line profile (as can be seen in \autoref{fig:spectrum}). \citet{WillsAndBrowne1986} found  (as part of a radio-loud AGN sample) an anti-correlation of the broad H$\beta$ line width with the ratio $R$ of the strengths of the radio core and lobes fluxes. Recent spectropolarimetric studies \citep{Aharonian2023} reveal that though 2021-2022, PKS 1510-089 remained in a low state of activity along with a decrease in optical polarization, resulting in optical spectra that can be explained through activity in the accretion disk and the BLR. During these periods \citet{Barnard2024} shows that the strength and width of the emission features (Mg II $\lambda2798\text{ \AA}$, H$\delta$, and H$\gamma$) remain unchanged, but with an average increase of the EW. In addition \citet{Podjed2024} could not observe the broad H$\gamma$ and H$\beta$ lines in polarized emission concluding that the emission line is intrinsically non-polarized.

The spectral energy distribution of PKS 1510-089 show the characteristic two hump morphology of blazars, with the low-energy peak attributed to synchrotron emission from the jet and thermal emission from the accretion disk, while the high-energy peak results from inverse Compton (IC) scattering of low-energy photons. Seed photons for IC scattering can originate from synchrotron radiation (SSC; \citealp{Maraschi1992}) or external sources like the accretion disk, BLR, or dusty torus (EC; \citealp{Sikora1994}). Several studies on the fitting of the SED (e.g. \citealp{Kataoka2008, DAmmando2009, Aleksic2014, Castignani2017}) model the high-energy peak with a combination of  SSC and EC, with one of them being the dominant contribution. In addition, the UV flux (responsable for the ionization of BRL material) has been traditionally modeled as a combination of synchrotron and thermal emission \citep{Bottcher2013}. However, \citet{Paliya2018} has model an additional IC contribution for the UV flux.

The M$_{BH}$ in PKS 1510-089 has been estimated to be in the order of $10^{8}\text{ M}_{\odot}$ using a variety of approaches. By analyzing the temperature profile of the accretion disk, \citet{Abdo2010} and \citet{Castignani2017} inferred the M$_{BH}$ to be  $5.40\times10^{8}\: M_{\odot}$ and $2.40\times10^{8}\: M_{\odot}$, respectively. Another approach, single-epoch spectra, was used by \citet{Oshlack2002} and \citet{Xie2005} using the $\lambda5100 \text{ \AA}$ continuum luminosity and the FWHM of H$\beta$ estimating a value of $3.86\times10^{8}\: M_{\odot}$, and $2.00\times10^{8}\: M_{\odot}$, respectively. \citet{Rakshit2020} used RM to calculate a mass of $5.71^{+0.62}_{-0.58}\times10^{7}\:M_{\odot}$.

In this study, we examine the variability of the H$\beta$ emission line, with its flux and FWHM, along with the optical continuum flux at $\lambda5100\text{ \AA}$. The spectra used were acquired from the Observatorio Astrofísico Guillermo Haro (OAGH), and the Steward Observatory (SO; \citealp{Smith2009C}). Our objective is to explore the relationship between the $L_{\lambda5100}$ and $L_{H\beta}$ luminosities and investigate the fluctuations spanning around 10 years of the FWHM$_{H\beta}$ against the $L_{\lambda5100}$ and $L_{H\beta}$ luminosities using correlation analysis. Furthermore, we separate the data sets based on the continuum dominant source, whether the accretion disk or jet, utilizing the non-thermal dominance (NTD) parameter \citep{Shaw2012, PatinoAlvarez2016}. This is to determine the role of the jet emission over the BLR across different activity epochs. Lastly, we determine the M$_{\text{BH}}$ through single-epoch spectra using $L_{H\beta}$ and FWHM$_{H\beta}$ for spectra where the accretion disk is the dominant ionization source for the BLR, avoiding contamination for synchrotron emission.

The cosmological parameters adopted throughout this paper are $H_{0}=71$ km s$^{-1}$ Mpc$^{-1}$, $\Omega_{\Lambda}=0.73$, and $\Omega_{M}=0.27$. At the redshift of the source, $z=0.361$, the luminosity distance is 1906.9 Mpc.

\section{Observations}\label{sec:obs}

We obtained optical spectra from two observatories: 34 spectra from the Observatorio Astrofísico Guillermo Haro (OAGH) and 353 spectra from the Steward Observatory (SO). The OAGH\footnote{\url{ https://astro.inaoep.mx/observatorios/oagh/}} spectra were observed under the spectroscopic monitoring program of bright $\gamma$-ray sources \citep{PatinoAlvarez2013Monitoring}. The spectra cover a wavelength range of $3800–7100\text{ \AA}$. We took spectra of a He-Ar lamp after each object exposure. This allows us to perform wavelength calibration and instrumental broadening correction. Table 1 of \hyperlink{paperI}{Paper I} shows the observation log of the spectra taken at OAGH. A 2.1m telescope with a Boller \& Chivens long-slit spectrograph\footnote{\url{ https://astro.inaoep.mx/observatorios/oagh/espectrografo-boller}} at the Cassegrain focus was used for the observations. This instrument achieved a spectral resolution of approximately $15\text{ \AA}$, with a grating of 150 l/mm. The spectroscopic data reduction was performed utilizing the \texttt{IRAF} package\footnote{\url{https://iraf-community.github.io}} \citep{Tody1986, Tody1993}, following standard procedures for bias and flat-field correction, cosmic-rays removal, 2D wavelength calibration, sky spectrum subtraction, and spectrophotometric calibration using standard stars that were observed each night.

The observations at the SO were carried out as part of the Ground-based Observational Support for the Fermi Gamma-ray Space Telescope at the University of Arizona monitoring program\footnote{\url{http://james.as.arizona.edu/~psmith/Fermi/}}. Utilizing the SPOL spectropolarimeter, the observations employed slit widths of $3\farcs0$, $4\farcs1$, and $5\farcs1$. The spectra were re-calibrated against the V-band magnitude and span over a decade, from 2008 to 2018, covering a wavelength range of $4000–7500\text{ \AA}$. For a detailed explanation of the observational setup and data processing, refer to \citet{Smith2009C}.

The process for fitting spectral features is thoroughly detailed in \hyperlink{paperI}{Paper I}. In our analysis, we apply corrections for both cosmological expansion to the flux, taking the form $(1+z)^{3}$ \citep{Peterson1997}, and galactic reddening using the dust maps from \citet{SchlaflyAndFinkbeiner2011}, with a color excess of $E(B-V)=0.09$ and a galactic reddening law with $R_{v}=3.1$ \citep{Cardelli1989}. The spectra were fitted with several components to measure the H$\beta$ and $\lambda5100\text{ \AA}$ continuum flux. The local continuum was characterized with a power-law function, the Fe II emission was modeled using the template from \citet{Kovacevic2010} that covers the wavelength range of $4000-5500\text{ \AA}$, and all emission and absorption lines were fitted with the aid of the \texttt{astropy.modeling}\footnote{\url{https://docs.astropy.org/en/stable/modeling/index.html}} framework. Three Gaussian functions were used to model the telluric absorptions near $5075\text{ \AA}$ in the rest frame. The H$\beta$ emission line was fitted using three Gaussians: a narrow, broad, and very broad component. The narrow component of the H$\beta$ line was constraining in its width and central wavelength based on the [O III]$\lambda5007\text{ \AA}$ line during profile fitting, allowing only its flux to vary. On the other hand, we allow to vary the central wavelength of the broad and very broad components. In consequence, the broad component in general is shifted less than $5\text{ \AA}$  from the narrow componen, while the very broad component is redshifted from the others to fit the characteristic red  ``bump" in PKS 1510-089 spectra. Each of the forbidden lines [O III]$\lambda\lambda4959,5007\text{ \AA}$ was modeled with a single Gaussian function.

\begin{figure}[b]
\centering
\includegraphics[width=\columnwidth]{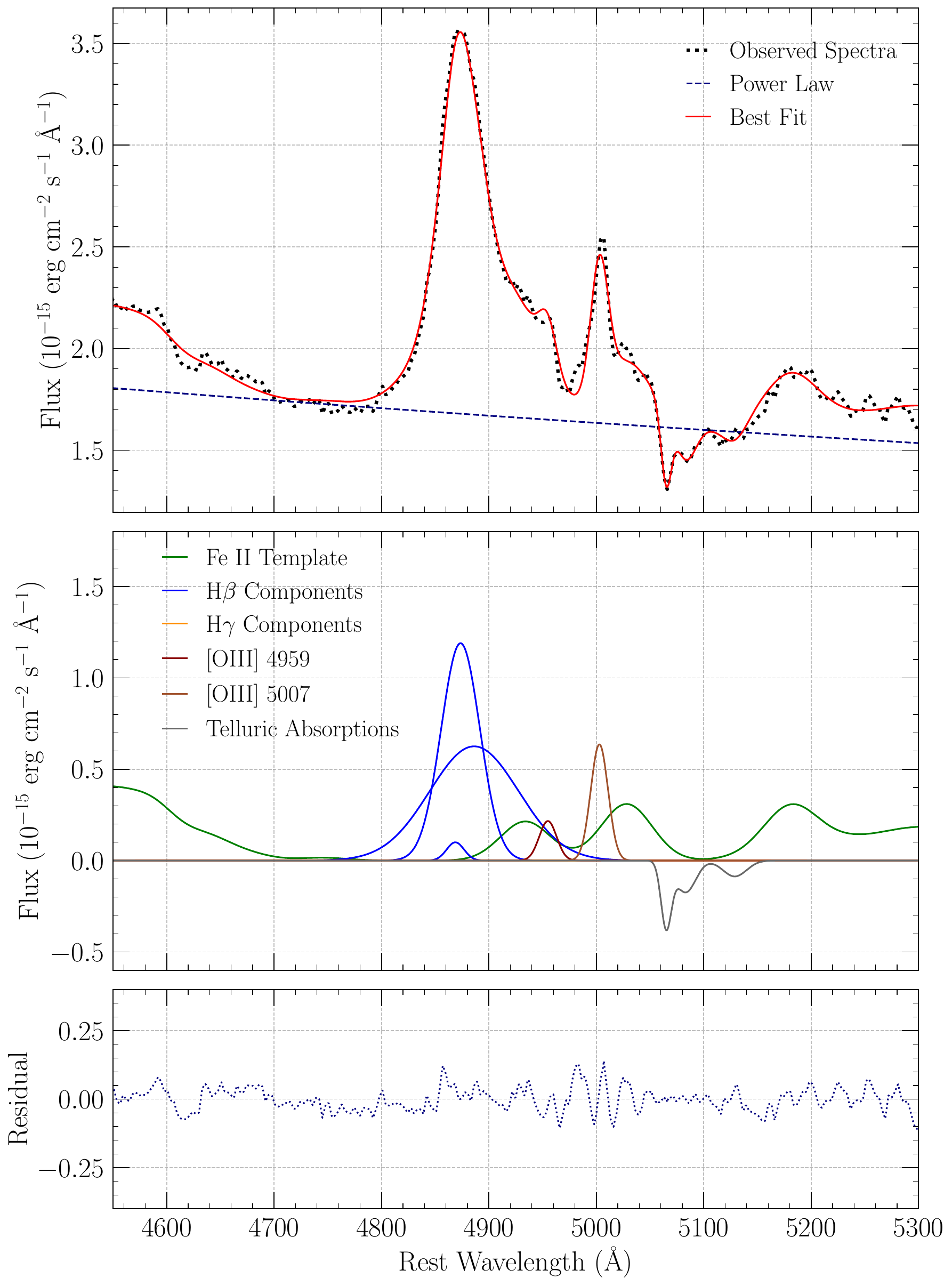}
\caption{Example decomposition of the H$\beta$ emission line from a spectrum observed at SO on March 19, 2010. Top panel: The rest-frame spectrum with the best-fit model overlaid. The continuum is represented by a power-law function. Middle panel: The broad and narrow components used to fit H$\beta$, along with the Fe II template, [O III] doublet, and telluric absorptions. Bottom panel: Residuals from the subtraction of the best-fit model from the observed spectrum.}
\label{fig:spectrum}
\end{figure}

Flux recalibration by [O III] emission lines was necessary to homogenize spectral measurements between observations made at the SO and the OAGH, explicit description is available on \hyperlink{paperI}{Paper I}. The $\lambda5100\text{ \AA}$ continuum flux was measured by first removing telluric absorption and Fe II emission from the spectrum, then calculating the mean flux value along with its uncertainty, which includes the standard deviation in the $5050-5150\text{ \AA}$ range and a flux calibration error, that is taken into account as the $10\%$ of the measured flux (Paul Smith private communication). The H$\beta$ flux was calculated by subtracting all components of the original spectrum, leaving only the respective line emission, and then integrating the remaining component in the range of $4781-5044\text{ \AA}$. Errors are calculated following the methodology described in \hyperlink{paperI}{Paper I} and references therein. In the H$\beta$ flux measurement, there are three unique contributions to the uncertainty, the first one is the random error caused by the dispersion and the signal-to-noise ratio (S/N) of the spectra. The second contribution is introduced by the subtraction of Fe II emission. The final contribution, resulting from flux calibration, is accounted for as $10\%$ of the measured flux. The uncertainty reported in the emission line flux calculation is derived from the quadratic sum of the three distinct sources of uncertainty. with median values of 2.15 (S/N ratio), 2.67 (iron), 13.1 (flux calibration), and 13.7 (total)$\times10^{-15}$ erg s$^{-1}$ cm${-2}$, respectively. An example of spectrum decomposition can be seen in  \autoref{fig:spectrum}. As shown in \hyperlink{paperI}{Paper I}, the narrow component of the H$\beta$ emission line is negligible with respect to the total line profile, representing from $5.2\%$ in its minimum to just $1.5\%$ in its maximum. It is noteworthy that this contribution is smaller than the uncertainty in the total H$\beta$ flux ($10.1\%$ and $17.1\%$ for minimum and maximum flux respectively). Therefore the total H$\beta$ profile is used for all calculations.

\begin{table*}[t]
\caption{Sample of flux measurements for the $\lambda5100\text{ \AA}$ continuum and H$\beta$ emission line, including the FWHM of H$\beta$.}
\label{tab:fluxes}
\begin{tabularx}{2\columnwidth}{lcccccccr}
\toprule
\multirow{2}{*}{JD$_{245}$} & \hspace{0.2in} & Flux $\lambda5100\text{ \AA}$ & \hspace{0.2in} & Flux H$\beta$ & \hspace{0.2in} & FWHM$_{H\beta}$ & \hspace{0.2in} & \multirow{2}{*}{Observatory} \\
& \hspace{0.2in} & ($\times10^{-15}$ erg s$^{-1}$ cm$^{-2}$ \AA\ $^{-1}$) & \hspace{0.2in} & ($\times10^{-14}$ erg s$^{-1}$ cm$^{-2}$) & \hspace{0.2in} & (km/s) & \hspace{0.2in} & \\
\hline
4830.03 & \hspace{0.2in} & $1.41\pm0.15$ & \hspace{0.2in} & $10.40\pm1.10$ & \hspace{0.2in} & $3098\pm218$ & \hspace{0.2in} & SO \\
4831.03 & \hspace{0.2in} & $1.17\pm0.12$ & \hspace{0.2in} & $9.26\pm0.97$ & \hspace{0.2in} & $3053\pm205$ & \hspace{0.2in} & SO \\
4832.02 & \hspace{0.2in} & $1.22\pm0.13$ & \hspace{0.2in} & $9.88\pm1.03$ & \hspace{0.2in} & $3209\pm226$ & \hspace{0.2in} & SO \\
4833.03 & \hspace{0.2in} & $1.27\pm0.13$ & \hspace{0.2in} & $8.94\pm0.94$ & \hspace{0.2in} & $3038\pm254$ & \hspace{0.2in} & SO \\
4860.02 & \hspace{0.2in} & $1.54\pm0.16$ & \hspace{0.2in} & $9.23\pm0.97$ & \hspace{0.2in} & $3250\pm207$ & \hspace{0.2in} & SO \\
4861.02 & \hspace{0.2in} & $1.42\pm0.15$ & \hspace{0.2in} & $8.86\pm0.92$ & \hspace{0.2in} & $3070\pm206$ & \hspace{0.2in} & SO \\
4862.03 & \hspace{0.2in} & $1.56\pm0.16$ & \hspace{0.2in} & $9.52\pm0.99$ & \hspace{0.2in} & $3147\pm234$ & \hspace{0.2in} & SO \\
4863.03 & \hspace{0.2in} & $1.54\pm0.16$ & \hspace{0.2in} & $9.99\pm1.03$ & \hspace{0.2in} & $3071\pm258$ & \hspace{0.2in} & SO \\
4864.02 & \hspace{0.2in} & $1.35\pm0.14$ & \hspace{0.2in} & $9.68\pm0.99$ & \hspace{0.2in} & $3194\pm211$ & \hspace{0.2in} & SO \\
4881.97 & \hspace{0.2in} & $1.23\pm0.13$ & \hspace{0.2in} & $9.69\pm1.01$ & \hspace{0.2in} & $3159\pm225$ & \hspace{0.2in} & SO \\
4882.97 & \hspace{0.2in} & $1.29\pm0.13$ & \hspace{0.2in} & $8.52\pm0.88$ & \hspace{0.2in} & $3247\pm200$ & \hspace{0.2in} & SO \\
\hline
\end{tabularx}
\begin{tablenotes}
\small
\item \textbf{Notes}. Hereinafter JD$_{245}$ represents JD-2450000. \\
(This table is available in its entirety in machine-readable form.)
\end{tablenotes}
\end{table*}

We were able to measure and analyze the evolution of the FWHM of the H$\beta$ emission line since the aforementioned Gaussians were fitted to the H$\beta$ profile. The uncertainty in the FWHM of the observed profile stems from two primary sources. The first is the inherent random error associated with the spectral dispersion, with a value of approximately $4\text{ \AA}$. The second source of uncertainty arises from the fitting of Gaussian functions to the profile. This uncertainty is quantified as the standard deviation between the fitted Gaussians and the observed spectrum, considering only flux values exceeding one-fifth of the peak H$\beta$ intensity. This approach focuses on the core of the profile, minimizing the impact of potential noise and fitting uncertainties in the wings. The observed flux uncertainty results from the quadratic sum of the two sources of uncertainty. The correction for instrumental broadening was made for each of the slit widths used in the SO observations. The instrumental broadening values for each slit width are taken from \citet{AmadorPortes2024a}. The spectra were observed with slit widths of $3\farcs0$, $4\farcs1$, and $5\farcs1$, with corresponding instrumental broadenings of $9.45\pm2.70\text{ \AA}$, $12.92\pm3.69\text{ \AA}$, and $16.07\pm4.59\text{ \AA}$ respectively. The uncertainty associated with the signal-to-noise ratio (S/N) was negligible compared to the large instrumental uncertainties for the different slit widths. The corrected profile was estimated as the quadratic subtraction of the observed and instrumental profiles, where its uncertainty was assessed from error propagation. A sample of the $\lambda5100\text{ \AA}$ continuum and H$\beta$ fluxes, as well as the FWHM of H$\beta$ are shown in \autoref{tab:fluxes}.

We computed the NTD parameter following the method outlined in \citet{Shaw2012}. This involved using the luminosities from the H$\beta$ emission line and the $\lambda5100\text{ \AA}$ continuum, as shown in \autoref{eq:NTD1}. Here, $L_{obs}$ represents the observed continuum luminosity, and $L_{pred}$ denotes the predicted continuum luminosity estimated from the H$\beta$ luminosity, based on a non-blazar sample relation \citep{GreeneAndHo2005}.

\begin{equation}
    NTD = \frac{L_{obs}}{L_{pred}}
\label{eq:NTD1}
\end{equation}

\begin{figure*}[t]
\centering
\includegraphics[width=2\columnwidth]{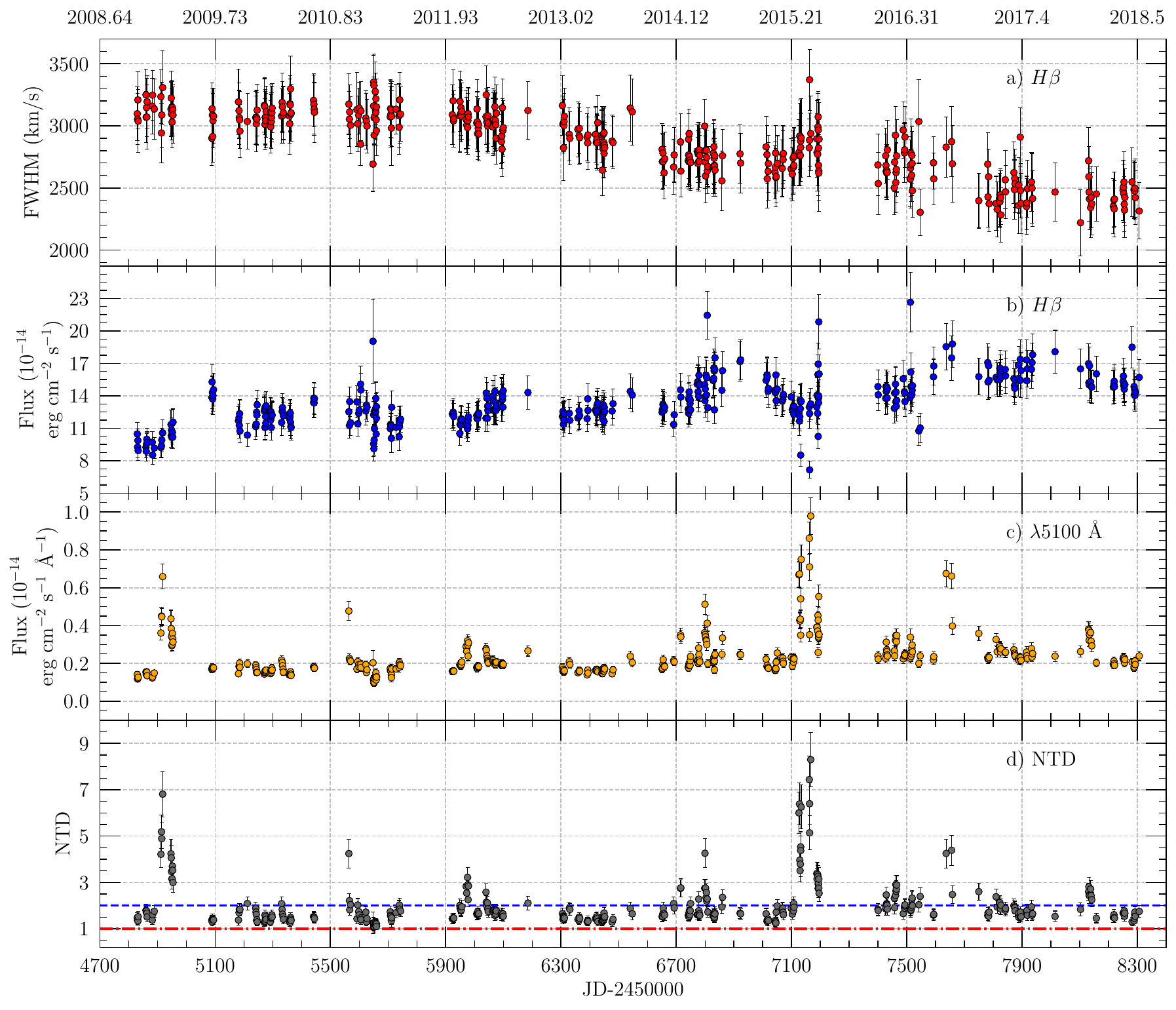}
\caption{Spectroscopic Light curves. (a) FWHM of the H$\beta$ emission line, (b) H$\beta$ emission line flux, (c) $\lambda5100\text{ \AA}$ continuum flux, and (d) NTD parameter. The red dash-dotted and blue dashed lines in panel (d) represent NTD$=1$ and NTD$=2$, respectively.}
\label{fig:speccurves}
\end{figure*}

The NTD parameter quantifies the relative contribution of the non-thermal emission (from the jet) compared to the total continuum emission in an AGN source. As detailed in \citet{PatinoAlvarez2016}, an NTD value of 1 indicates purely thermal emission from the accretion disk. For $1<$ NTD $<2$, the disk remains dominant, but the jet contributes. At NTD $= 2$, both components contribute equally. When NTD $>2$, the jet dominates the continuum emission. Consequently, the observations can be categorized into two regimes: Jet dominance (NTD $>2$) and Disk dominance (NTD $< 2$). The light curves of the H$\beta$ and $\lambda 5100\text{ \AA}$ continuum fluxes along with the FWHM of $H\beta$ and NTD are displayed in \autoref{fig:speccurves}.

\section{Correlation Analysis}\label{sec:correlation}

In \hyperlink{paperI}{Paper I}, we find a delay consistent with zero between optical/IR emissions ($\lambda5100\text{ \AA}$ continuum, J-band, V-band, and NTD), indicating nearly simultaneous emissions from co-spatial regions. High NTD values (NTD $>2$) are associated with jet-dominated continuum emission, indicating that the variability in the $\lambda5100\text{ \AA}$ continuum, J-band, and V-band is primarily driven by synchrotron emission from the jet. Furthermore, we observed a delay of approximately $80\pm6$ days between the $\lambda5100\text{ \AA}$ continuum and the H$\beta$ emission line flux. This delay is interpreted as the distance between the continuum emission source and the BLR. Since most of the continuum variability comes from the jet emission, we expect that this delay mostly traces the variability of the jet during high activity states. Therefore we look for correlations in the logarithmic space between FWHM$_{H\beta}$, $L_{\lambda5100}$, and $L_{H\beta}$ for the entire dataset, as well as separately for the jet dominance and disk dominance regimes to picture the role of jet emission over the H$\beta$ emission line (and thus the BLR).

\subsection{Luminosity Relations}\label{sec:Lrelations}

The NTD parameter is the ratio of observed to predicted continuum luminosity for an AGN source. In this case, the predicted luminosity is estimated from the H$\beta$ emission line using a relationship derived from a non-blazar sample by \citet{GreeneAndHo2005}. Solving Equation 2 from \citet{GreeneAndHo2005} yields $L_{\lambda5100}$ along with its respective error propagation, defining the relationship given by \autoref{eq:LumLum}, where $L_{\lambda5100}\equiv L_{pred}$.

\begin{figure*}[t]
\centering
\includegraphics[width=2\columnwidth]{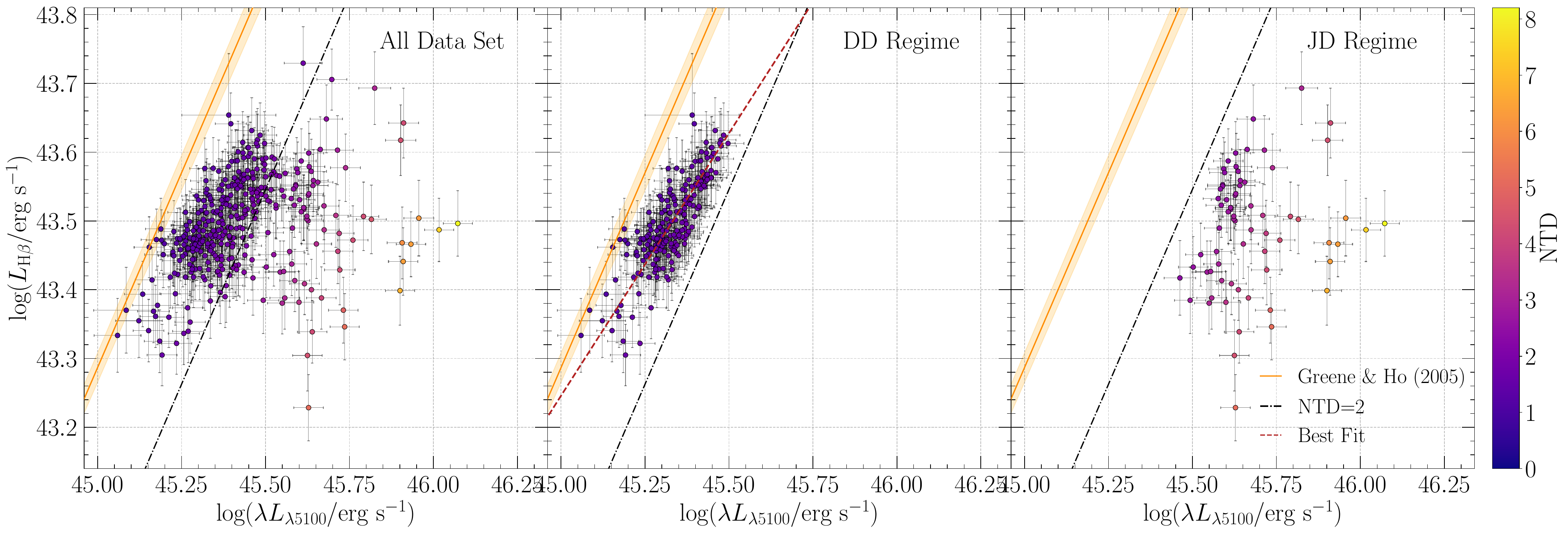}
\caption{Variation of the H$\beta$ emission-line luminosity compared to the $\lambda5100\text{ \AA}$ continuum luminosity. In all panels, the color bar indicates the NTD value for each observation. Left Panel: Full sample. Middle Panel: Disk dominance (DD) regime data. Right Panel: Jet dominance (JD) regime data. The dashed black line denotes the boundary between the regimes. The orange solid line and shaded area represent the \citet{GreeneAndHo2005} relation for a non-blazar sample and its uncertainty at $3\sigma$. The dashed red line denotes the significant ($p_{v}<0.05$) linear regression to the correspondent data.}
\label{fig:LLrelation}
\end{figure*}

\begin{table}[b]
\centering
\caption{Pearson ($\rho_{P}$) and Spearman ($\rho_{S}$) correlation coefficients obtained for L$_{\lambda5100}$ versus L$_{H\beta}$ across various data sets: Full Set, Disk Dominance (DD), and Jet Dominance (JD). Corresponding p-values are displayed for each case.}
\label{tab:correlations1}
\begin{tabularx}{\columnwidth}{lcccccr}
\toprule
Regime &  & $\rho_{P}$ & $p_{v}$ & $\rho_{S}$ & & $p_{v}$ \\
\hline
Full Set & & -0.200 & $2.60\times10^{-6}$ & -0.450 & & $6.0\times10^{-21}$ \\
DD & & -0.800 & $6.90\times10^{-59}$ & -0.810 & & $7.00\times10^{-53}$ \\
JD & & -0.100 & 0.260 & -0.170 & & 0.200 \\
\hline
\end{tabularx}
\end{table}

We compare their luminosities, relating them to \autoref{eq:LumLum}. \autoref{fig:LLrelation} illustrates the luminosity relationship, marking the boundary between the jet dominance and disk dominance regimes at NTD $=2$. We also performed a linear regression analysis using orthogonal distance regression from the \texttt{SciPy ODR package}. However, the fit turned out to not accurately represent the data, as evidenced by the p-value ($p_{v}$) of 1 (to machine accuracy). The p-value represents the probability that our null hypothesis (which states there is no relationship between the model and the data) is true. A p-value below 0.05 is considered statistically significant, indicating that the observed relationship is unlikely to have occurred by chance. Additionally, we conducted Pearson ($\rho_{P}$) and Spearman ($\rho_{S}$) correlation rank tests, as displayed in \autoref{tab:correlations1}. Person or Spearman coefficients with values absolute below $|\rho|\leq0.39$ are consider as a weak correlation,  values between $0.40\leq|\rho|\leq0.59$  as a moderate correlation, and values greater than $0.60\leq|\rho|$  as a strong correlation (positive or negative, given the case). The results show a weak-to-moderate correlation between the continuum and H$\beta$ emission line luminosities.

\begin{equation}
\begin{split}
L_{\lambda5100}=&10^{44}\text{ erg s}^{-1}\\
&\left(\frac{L_{H\beta}}{(1.425\pm0.007)\times10^{42}\text{ erg s}^{-1}}\right)^{-\frac{1}{1.133\pm0.005}}
\end{split}
\label{eq:LumLum}
\end{equation}

Furthermore, we divided the sample into jet dominance and disk dominance regimes considering uncertainties: for the disk dominance regime, only data points with NTD$+\sigma<2$ were included, whereas, for the jet dominance regime, data points with NTD$-\sigma>2$ were used. This separation ensures that each subsample predominantly captures emission from either the jet or the accretion disk, minimizing contamination from spectra near the threshold. In \autoref{fig:LLrelation} the disk dominance regime dataset consists of 219 points, accounting for $56.6\%$ of total observations. Conversely, $17.8\%$ of the spectra indicate the jet as the dominant source of the continuum emission. The remaining $25.6\%$ of observations have NTD values around 2 within $1\sigma$ uncertainty, making their regime classification uncertain.

For the disk dominance regime, the linear regression yields a slope of $\beta=0.76\pm0.05$, with a p-value of 0 (to machine accuracy). Moreover, both the Pearson correlation tests show a strong correlation. This suggests that during periods when the accretion disk dominates the continuum emission, the behavior of the emission line relative to the continuum exhibits a strong positive correlation, as observed in the non-blazar AGN sample studied by \citet[Equation 2]{GreeneAndHo2005}. Conversely, in the jet dominance regime, the slope of the linear fit was not reliable ($p_{v}=1$). The Pearson correlation, and Spearman correlation tests indicate no significant correlation between the continuum and emission line luminosities during jet dominance periods. Specific correlation values are summarized in \autoref{tab:correlations1}.

\subsection{Luminosity and FWHM Correspondence}\label{sec:FWHMandL}

In \autoref{sec:obs}, we corrected the measured emission line widths for instrumental broadening, enabling a comparison with luminosity values to explore the behavior of the H$\beta$ emission line FWHM across different regimes of continuum dominance. We analyzed correlations between FWHM$_{H\beta}$, $L_{\lambda5100}$, and $L_{H\beta}$ for the entire dataset, as well as separately for the jet dominance and disk dominance regimes.  Correlation coefficients, $\rho_{P}$ and $\rho_{S}$, were calculated for each case, along with their respective p-values, as summarized in \autoref{tab:correlations}. In addition, a lineal regression analysis was also made for each case.

\begin{table}[t]
\centering
\caption{Pearson ($\rho_{P}$) and Spearman ($\rho_{S}$) correlation coefficients obtained for L$_{H\beta}$ or L$_{\lambda5100}$ versus FWHM$_{H\beta}$ across various data sets: Full Set, Disk Dominance (DD), and Jet Dominance (JD). Corresponding p-values are displayed for each case.}
\label{tab:correlations}
\begin{tabularx}{\columnwidth}{lcccccr}
\toprule
Regime &  & $\rho_{P}$ & $p_{v}$ & $\rho_{S}$ & & $p_{v}$ \\
\hline
& & \multicolumn{3}{c}{L$_{\lambda5100}$ vs FWHM$_{H\beta}$} & & \\
\hline
Full Set & & -0.187 & $2.09\times10^{-4}$ & -0.450 & & $2.27\times10^{-20}$ \\
DD & & -0.678 & $8.81\times10^{-31}$ & -0.650 & & $6.21\times10^{-28}$ \\
JD & & -0.104 & 0.395 & -0.054 & & 0.658 \\
\hline
& & \multicolumn{3}{c}{L$_{H\beta}$ vs FWHM$_{H\beta}$} & \\
\hline
Full Set & & -0.650 & $9.34\times10^{-48}$ & -0.683 & & $1.72\times10^{-54}$ \\
DD & & -0.678 & $8.04\times10^{-31}$ & -0.651 & & $7.86\times10^{-28}$ \\
JD & & -0.575 & $2.33\times10^{-7}$ & -0.630 & &  $6.59\times10^{-9}$ \\
\hline
\end{tabularx}
\end{table}

\begin{figure*}[t]
\centering
\includegraphics[width=2.05\columnwidth]{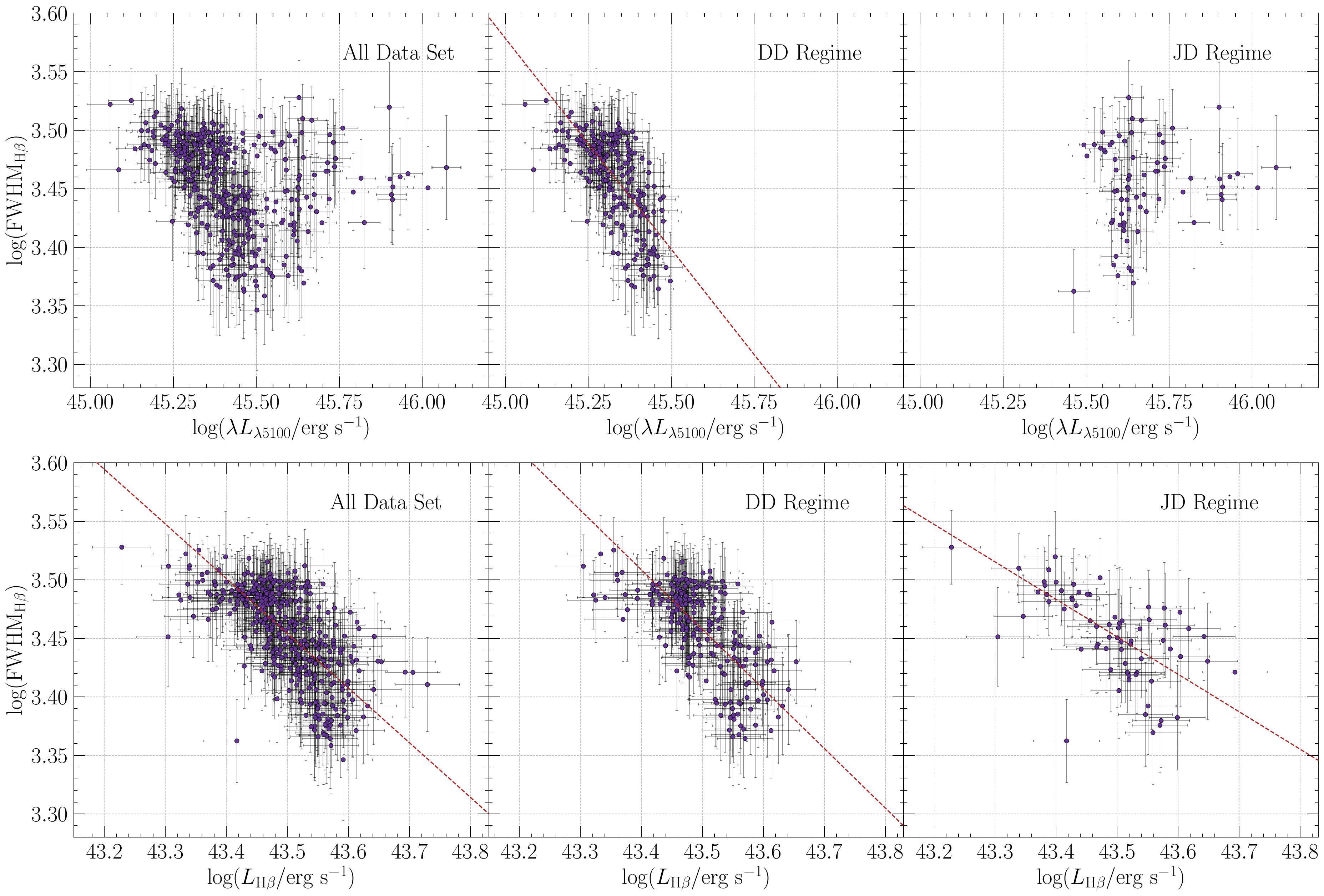}
\caption{Top Row: Variation of the H$\beta$ emission-line FWHM compared to the $\lambda5100\text{ \AA}$ continuum luminosity. Bottom Row: Variation of the H$\beta$ emission-line FWHM compared to its luminosity. Left Column: Full sample. Middle Column: Disk dominance (DD) regime data set. Right Column: Jet dominance (JD) regime data set. Only lineal regression fits with $p_{v}<0.05$ are plotted, as red dashed lines, for the corresponding data.}
\label{fig:LFWHMrelations}
\end{figure*}

The Pearson correlation test revealed no correlation between $L_{\lambda5100}$ and FWHM$_{H\beta}$ (se \autoref{fig:LFWHMrelations}, Top Row, Left Panel) across the entire dataset. In contrast, the Spearman correlation test reveals  a moderate anti-correlation, indicating a monotonic, though not linear, relationship between $L_{\lambda5100}$ and FWHM$_{H\beta}$.  The linear regression model was not statistically significant ($p_{v}=0.620$), rendering the estimated slope meaningless. Strong correlation values were observed in the disk dominance regime for both correlation tests, while this moderate, statistically significant anti-correlation is absent in the jet dominance regime.

In the disk dominance regime, where $L_{\lambda5100}$ primarily originates from the accretion disk, serving as the ionization source for the BLR, an increase in ionizing photons (and $L_{\lambda5100}$) expands the BLR radius, ionizing clouds further from the central engine, a phenomenon known as ``breathing-BLR'' (e.g. \citealp{Peterson1992, Peterson2002, Barth2015}). This process manifests as a narrowing of the H$\beta$ line profile, consistent with the observed distribution of data points. Conversely, in the jet dominance regime, the lack of correlation suggests that jet emission has minimal impact on the distribution of FWHM$_{H\beta}$. The ``breathing" behavior observed in the disk dominance regime indicates that the BLR adapts its size and velocity structure in response to fluctuations in ionizing flux from the accretion disk. 

In contrast, the absence of ``breathing" during the jet dominance regime, implies that the ionizing influence of the jet on the BLR is weaker compared to the accretion disk, thus insufficient to induce a similar dynamic response in the BLR. For the entire dataset and jet dominance regime, the high p-values of the linear regression fits, show that the data cannot be accurately described by a linear model. On the other hand, in the disk dominance regime we found a slope value of $\beta=-0.360\pm0.034$ (p-value $=0.0$, to machine accuracy).

When examining the properties of the H$\beta$ emission line, we found a strong anti-correlation between $L_{H\beta}$ and FWHM$_{H\beta}$ across all data subsets, as illustrated in the bottom row of \autoref{fig:LFWHMrelations}. This suggests a scenario where changes in the H$\beta$ luminosity affect the BLR similarly regardless of the primary ionizing source. This effect can also be observed as a long-term trend in the light curves of H$\beta$ (FWHM and flux, \autoref{fig:speccurves}a,b), where an increase in flux corresponds to a decrease in FWHM over the approximately 10 years that cover our observations data. The linear regression fit was statistically significant for all the subsets; the entire data set, disk, and jet dominance regime, yielding a slope value of $\beta=-0.467\pm0.032$ (p-value $=0.00$, to machine accuracy), $\beta=-0.509\pm0.048$ (p-value $=0.00$, to machine accuracy), and $\beta=-0.320\pm0.056$ (p-value $=0.002$) respectively.

\section{Black Hole Mass Estimation}\label{sec:SMBHmass}

The variability in this source is predominantly driven by the jet, as indicated by the NTD parameter. This jet dominance complicates the estimation of the supermassive black hole mass ($M_{BH}$) via reverberation mapping, which relies on the delay between an emission line and the continuum to approximate the BLR radius ($R_{BLR}\approx c\tau$). When the continuum variability is jet-dominated, the measured delay only represents the distance between the BLR and the jet, rather than the true size of a virialized BLR. In contrast, for data within the disk dominance regime, where the accretion disk is the primary source of continuum emission, the delay is expected to accurately reflect the BLR radius.

However, as discussed in \hyperlink{paperI}{Paper I}, the cross-correlation analysis between the $\lambda5100\text{ \AA}$ continuum and H$\beta$ emission fluxes in the disk dominance regime yielded inconclusive delay results, thereby hindering reliable reverberation mapping. Given the challenges with reverberation mapping, we opted to use spectroscopic single-epoch methods for $M_{BH}$ estimation, focusing exclusively on observations within the disk dominance regime. Specifically, we applied \autoref{eq:SMBH} from \citet{GreeneAndHo2005} to the 192 spectra in the disk dominance regime. This approach leverages the corrected emission line widths and luminosity values to estimate $M_{BH}$, ensuring that jet-dominated variability does not influence the results.

\begin{equation}
\begin{split}
M_{\text{BH}}=(3.6\pm0.2)\times10^{6}&\left(\frac{L_{H\beta}}{10^{42}\text{ erg/s}}\right)^{0.56\pm0.02}\\
&\left(\frac{FWHM_{H\beta}}{10^{3}\text{ km/s}}\right)^{2}\:M_{\odot}
\label{eq:SMBH}
\end{split}
\end{equation}

As previously discussed, FWHM$_{H\beta}$ is corrected for instrumental broadening. From this set of data points (middle bottom panel of \autoref{fig:LFWHMrelations}), we estimated a weighted mean mass for the supermassive black hole ($M_{BH}$). The resulting value is $M_{BH}=2.85\pm0.37\times10^{8}\:M_{\odot}$, with the uncertainty derived from the standard deviation of this set.

\section{Discussion}\label{sec:discussion}
\subsection{Differences in Spectral Features for Different Regimes}\label{sec:spectralfeat}

Based on the H$\beta$ and $\lambda5100\text{ \AA}$ continuum light curves shown in \autoref{fig:speccurves} and the approximately $80$-day delay identified in \hyperlink{paperI}{Paper I}, we observe that flaring events in the $\lambda5100\text{ \AA}$ continuum are followed by corresponding increases in the H$\beta$ emission line flux. This, along with the high NTD values associated with these events, suggests an interference of jet emission with the H$\beta$ emission line. Additionally, \autoref{fig:LLrelation} shows that the highest values of $L_{H\beta}$ occur in the jet dominance regime or at the boundary between regimes.

An anti-correlation was also found in the logarithmic space when comparing the FWHM$_{H\beta}$ against $L_{\lambda5100}$, but only in the disk dominance regime. This implies a scenario reminiscent of a ``breathing-BLR''. This phenomenon describes how the BLR surrounding the SMBH responds dynamically to fluctuations in the ionizing continuum flux originating from the accretion disk. As the emission from the accretion disk intensifies or diminishes, the BLR undergoes a corresponding expansion or contraction. This behavior occurs because changes in the ionizing photon flux directly impact the extent of ionized material.  The absence of this anti-correlation in the JD regime can be attributed to jet contamination. Unlike the accretion disk, the jet does not induce the same ``breathing'' effect in the BLR given the non-isotropic nature of synchrotron emission. The ``breathing'' effect underscores the responsiveness of the BLR to changes in the ionizing continuum flux from the accretion disk, a key aspect of its variability and structure.

Moreover, the detection of the ``breathing-BLR'' effect when comparing FWHM$_{H\beta}$ with $L_{H\beta}$ across different datasets suggests that the photons responsible for H$\beta$ emission originate from the canonical BLR, irrespective of whether the primary continuum source is the accretion disk or the jet. This implies that the emission is not from an external BLR material. Given these observations, we propose that the base of the jet (where the UV synchrotron emission is produced) is embedded within the canonical BLR. This scenario suggests that jet emission can influence the H$\beta$ emission line, which is typically emitted from the canonical BLR. Consequently, it appears that dual ionization sources are affecting the BLR clouds, with varying levels of jet contamination present at all times. Given this scenario for the jet-BLR it is crucial to identify the mechanism producing ionizing UV photons: synchrotron emission or IC scattering. SED fitting by \citet{Bottcher2013} suggests synchrotron emission for UV photons, with significant thermal contribution from the accretion disk. However, \citet{Paliya2018} models indicate a regime shift around $10^{16}$ Hz ($\sim300\text{ \AA}$). This implies that photons between $900\text{ \AA}$ (responsible for H ionization) and $300\text{ \AA}$ are primarily synchrotron-dominated, while those below $300\text{ \AA}$ are dominated by SSC scattering.

\begin{figure}[t]
\centering
\includegraphics[width=\columnwidth]{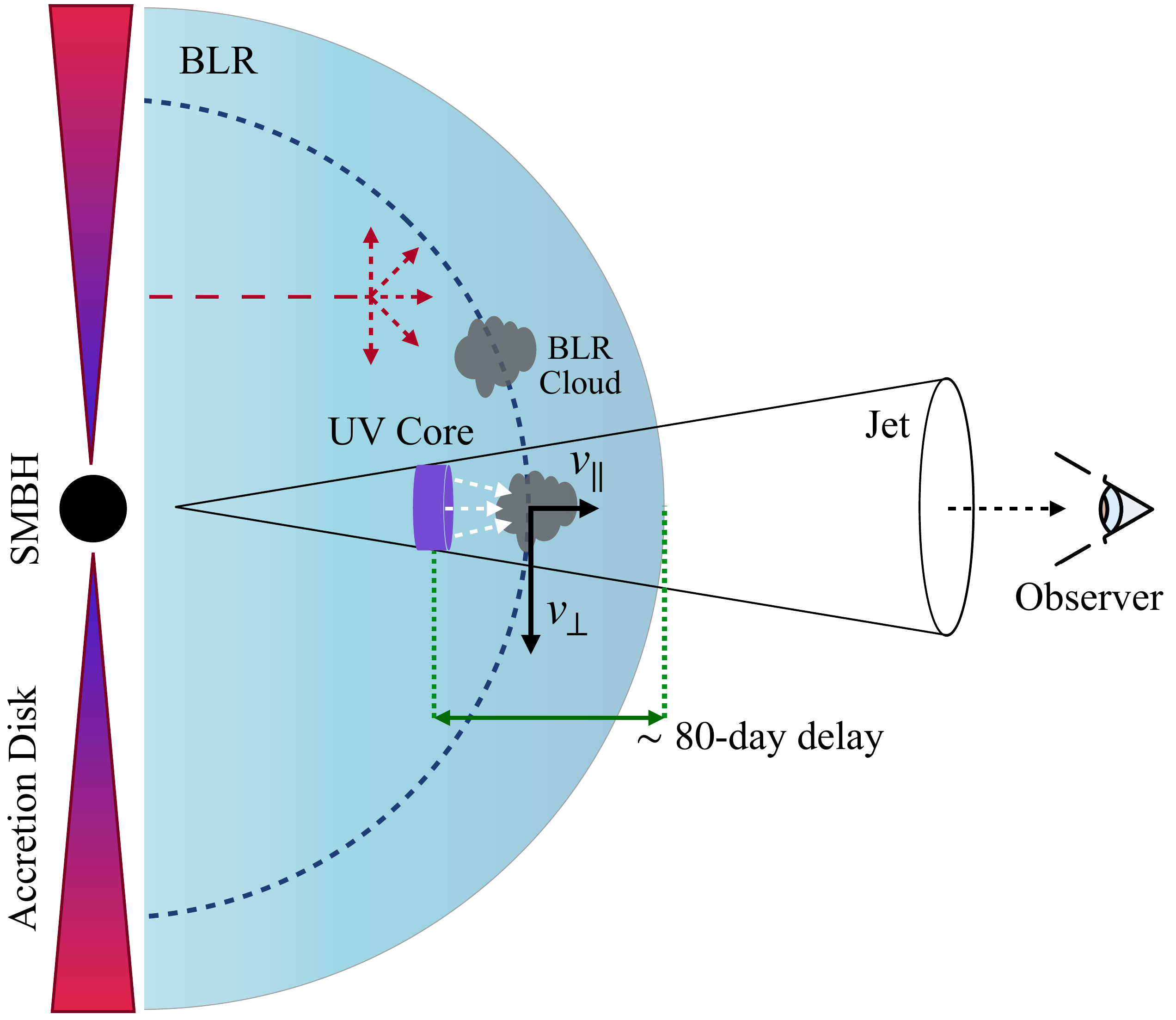}
\caption{Schematic illustration of the Jet-BLR system suggested in this work (not to scale). The blue dashed line represent the orbit of a BLR cloud around the SMBH with the parallel velocity component being low while crossing the jet section. BLR clouds in this region will be ionized by a contribution of UV flux from the accretion disk (red arrows) and from the jet (white arrows). The $\sim80$-day delay between the continuum emission and the H$\beta$ emission line found in \citet{AmadorPortes2024b} trace the distance between the continuum emission within the jet during flare-like events and the edge of the BLR.}
\label{fig:schematic}
\end{figure}

During flare-like events, the jet emerges as the primary source of continuum emission, as evidenced by the NTD analysis, and the increases in emission line flux during these. Combining this with our previous findings provides valuable insights into the dynamics of the BLR-Jet system. While the base of the jet resides within the BLR, it is important to note that the ionizing emission of the jet is not isotropic due to its collimated nature. As a result, it likely impacts only a fraction rather than the entirety of the BLR, with the opening angle of the jet-cone being the parameter that determines the fraction of the canonical BLR that will be ionized. The variability in the jet emission further complicates matters by disrupting the typical  ``breathing'' relationship observed when the accretion disk dominates. Unlike the relatively steady flux from the accretion disk, the variable nature of the jet emission can lead to irregular responses in the BLR, affecting its size and velocity structure in ways that differ from accretion-driven variations. Specially, since the jet is pointed towards the observer, the expectation for the orbit around the SMBH of a BLR cloud that is passing through the jet-cone, is that its radial velocity is very low, since the movement of the cloud is almost perpendicular to the jet direction in that specific region. This selective interaction is likely responsible for the observed approximately $80$-day delay, particularly prominent during periods of heightened activity, since the jet emission is persistent, the measured delay reflects the distance between the edge of the BLR and the region within the jet where continuum emission originates during flare-like events. This delay, therefore, does not correspond to the size the BLR. In \autoref{fig:schematic} we present a schematic illustration of the Jet-BLR system coupling and the different interactions between regions.

Although the jet can ionize portions of the BLR, it is unlikely to completely ``clean'' the BLR region due to several factors. First, the aperture of the jet angle in the central parsecs is extremely narrow, approximately $0.2\pm0.2$ degrees, as reported by \citet{Jorstad2005b}, resulting in a confined cone of influence along the jet direction. If the BLR has a spherical or toroidal geometry, most of its material would lie outside the path of the jet, avoiding direct interaction with the jet plasma. Additionally, the distribution of the BLR, with gas clouds over scales of light days to light months \citep{Peterson1997}, with different densities between them, further limits the regions directly affected by the jet, leaving significant portions of the BLR intact. The ability of the jet to remove BLR material also depends on the density of the interacting regions. While the jet is highly energetic, the BLR clouds have very high densities (e.g., $n_{e}\sim10^{9}–10^{11} \text{ cm}^{-3}$; \citealp{Blandford1990}), which could withstand the force exerted by the jet. As shown in \autoref{fig:schematic}, if the base of the jet is embedded within the BLR, a specific distance must be maintained between the base and the SMBH. Consequently, layers of the BLR below the jet base would remain unaffected by their activity, with only the outer layers of the BLR experiencing any impact from the jet. Furthermore, as the jet entrains BLR material, its kinetic energy may dissipate, reducing its ability to fully clear the region. Finally, jet activity is episodic rather than continuous, characterized by quiescent phases interspersed with flaring events. During these quiescent phases, the BLR can replenish itself through inflows of gas from the accretion disk or external sources, further ensuring its persistence despite occasional jet interactions

A future analysis to test the proposed scenario is with the asymmetry parameters (asymmetry, asymmetry index, and kurtosis). It is possible that if some clouds are displaced in the direction of the jet, an anisotropy would be observed on the H$\beta$ emission line profile. For example, asymmetric profiles might arise due to preferential ionization of BLR clouds near the jet axis. Complementary to this, a RMS spectra could show an asymmetry on the blue side, given the close alignment between the jet and our line of sight. However, these tests are beyond the scope of this paper.

The luminosity relation between the $\lambda5100\text{ \AA}$ continuum and H$\beta$ observed across our entire data set demonstrates a weak correlation, particularly diverging from the established relation for non-blazar samples \citep{GreeneAndHo2005} when NTD exceeds 2. This pattern is consistent with findings by \citet{Rakshit2020}. Similar behaviors have been noted in other sources such as CTA 102 \citep{Chavushyan2020} and 3C 454.3 \citep{AmayaAlmazan2021}, where correlations between $\lambda3000\text{ \AA}$ continuum and Mg II $\lambda2798\text{ \AA}$ fluxes were examined, as well as in B2 1633+382 \citep{AmayaAlmazan2022}, involving $\lambda1350\text{ \AA}$ continuum and C IV $\lambda1549\text{ \AA}$ fluxes. However, when focusing exclusively on the disk dominance regime, we observed a strong positive correlation similar to that seen in the non-blazar sample. This suggests that when the accretion disk is the primary source of emission, the luminosity relationship mirrors the behavior observed in non-blazar AGNs.

\subsection{Black Hole Mass Measurements}\label{masscompar}

As previously discussed, the variability of PKS 1510-089 is primarily attributed to its jet activity. Consequently, the delay observed between the $\lambda5100\text{ \AA}$ continuum and the H$\beta$ emission line flux does not correspond to the size of the BLR but rather indicates the distance between the BLR and the jet. This precludes the application of reverberation mapping techniques unless a delay observed in the disk dominance regime is identified, which was not detected in our study. An alternative approach involves using single-epoch spectra from periods with minimal jet contamination, specifically spectra with NTD$+\sigma<2$. This selection criterion helps mitigate the influence of jet emissions. Subsequently, we calculated the $M_{BH}$ for each spectrum using the scaling relation derived from FWHM$_{H\beta}$ and $L_{H\beta}$ by \citet{GreeneAndHo2005}. A weighted mean value of $M_{BH}=2.85\pm0.37\times10^{8}\:M_{\odot}$ was determined. We also estimated the mass for the entire spectra set to quantify the jet contamination over the $M_{BH}$. For this complete set, we obtain a value of  $M_{BH}=2.80\pm0.37\times10^{8}\:M_{\odot}$. The result is consistent within $1\sigma$ with the reported value utilizing spectra in the disk dominance regime. This advises, that for this object, using the complete spectra set to estimate the $M_{BH}$ does not bias the obtained value. This supports the proposed scenario in which the base of the jet is embedded within the BLR but affects only a fraction of it due to the non-isotropic nature of jet synchrotron emission.

Our result aligns well with previous estimations. \citet{Oshlack2002} determined the black hole mass of PKS 1510-089 to be $3.86\times10^{8}\: M_{\odot}$ using single-epoch spectra, which is only approximately 1.35 times larger than our finding. Similarly, \citet{LiangAndLiu2003} and \citet{Xie2005} estimated black hole masses of approximately $1.58\times10^{8}\: M_{\odot}$ and $2.00\times10^{8}\: M_{\odot}$, respectively, based on minimum variability time scales and single-epoch spectra. Notably, these techniques heavily depend on the choice of scaling relation. By exclusively utilizing spectra from the disk dominance regime spanning approximately 10 years, we achieved a robust estimation of the black hole mass with minimal jet contamination and short-term variability potentially related to the jet. \citet{Abdo2010} and \citet{Castignani2017} employed accretion disk modeling with UV data to estimate $M_{BH}$, obtaining values of $5.40\times10^{8}\: M_{\odot}$ and $2.40\times10^{8}\: M_{\odot}$, respectively. Additionally, \citet{Rakshit2020} applied the reverberation mapping technique and reported a mass of $5.71^{+0.62}_{-0.58}\times10^{7}\:M_{\odot}$, where NTD parameter was primarily, but not exclusively, below 2 (Figure 2 in \citealp{Rakshit2020}). This discrepancy may explain our inability to detect a delay in the disk dominance regime. It needs to be considered that the calculation of the NTD values relies on the chosen luminosity relation, where the $L_{\lambda5100}-L_{H\beta}$ relation used in their analysis was retrieved from \citet{Rakshit2020b}.

Consequently, a broad range of black hole masses, spanning from $1.5-5.5\times10^{8}\: M_{\odot}$ (excluding the value of approximately $10^{7}\: M_{\odot}$), has been reported in the literature for PKS 1510-089. This diversity suggests methodological robustness across different approaches to estimating $M_{BH}$. Our estimation is derived from directly observable variables in a large spectrum set, aiming to minimize the influence of jet-related effects and short-term variability.

\section{Summary}

We studied the blazar PKS 1510-089, known for its flares across various wavelengths, using 10 years of spectroscopic light curve data. Our research utilized optical spectroscopic data from the Observatorio Astrofísico Guillermo Haro and the Steward Observatory enabling us to characterize the profile of the prominent H$\beta$ emission line from which we obtain their flux and FWHM, as well as the $\lambda5100\text{ \AA}$ continuum. We perform a correlation analysis between the $L_{\lambda5100}$, $L_{H\beta}$, and FWHM$_{H\beta}$, yielding us to the following key findings.

\begin{enumerate}
    \item The luminosity relation in the logarithmic space between the $\lambda5100\text{ \AA}$ continuum and H$\beta$ observed in our entire dataset reveals a weak positive correlation, with the data diverging from the relation established for a non-blazar sample \citep{GreeneAndHo2005}, at NTD$>2$. This pattern has also been found in other FSRQs sources for the  $L_{\lambda3000}-L_{\text{Mg II}}$ and $L_{\lambda1350}-L_{\text{C IV}}$ relations. The correlation intensifies in the disk dominance regime with a Pearson and Spearman correlation rank test of $0.80$ and $0.79$ respectively. Hence, the behavior of the emission line relative to the continuum emission follows a positive relation like the one derived from non-blazar AGN.
    \item We found a consistent anti-correlation between $L_{H\beta}$ and FWHM$_{H\beta}$ across all data sets (full set, disk dominance, and jet dominance regimes) in the logarithmic space. Jet-induced ionization does not significantly affect the FWHM$_{H\beta}$ distribution, suggesting that the BLR responds similarly to changes in ionizing continuum flux, regardless of the primary ionizing source. Furthermore, an anti-correlation observed when comparing against $L_{\lambda5100}$ suggests a ``breathing-BLR" effect. This anti-correlation is absent in the jet dominance regime, likely due to the anisotropy of the jet continuum emission. Our findings support the existence of dual ionization sources within the BLR, where the accretion disk and the jet contribute differently to its ionization and dynamical behavior. We hypothesize that the base of the jet is embedded within the BLR.
    \item  In \hyperlink{paperI}{Paper I}, cross-correlations across the full dataset between the $\lambda5100\text{ \AA}$ continuum and H$\beta$ emission line fluxes revealed a delay of approximately $80\pm6$ days. Therefore, given the scenario of the persistent jet emission, this delay signifies the separation between the BLR edge and the continuum emission region within the jet during flare-like events. Thus this delay cannot be interpreted as the size of a virialized BLR. 
    \item Single-epoch spectra were employed to estimate $M_{BH}$ in the disk dominance regime, using the scaling relation proposed by \citet{GreeneAndHo2005}. A robust $M_{BH}$ value of $2.85\pm0.37\times10^{8}\:M_{\odot}$ was derived exclusively from spectra spanning approximately a decade within the disk dominance regime. This approach minimized jet contamination and short-term variability potentially associated with jets. However, for this object, the jet does not significantly impact the $M_{BH}$ estimated, as seen from the analysis of the complete spectra set for this particular source.
\end{enumerate}

In summary, our findings underscore the complex interplay between the accretion disk and the jet in shaping the conditions of the BLR structure. The non-isotropic nature, and variability of the jet emission, introduce significant deviations from expectations under accretion-dominated conditions, highlighting the need for nuanced interpretations of BLR dynamics in active galactic nuclei.

\section*{Acknowledgments}
We thank the anonymous referee for the constructive comments that helped to improve the manuscript. A. A.-P. gratefully acknowledges the support received from the CONAHCYT (Consejo Nacional de Humanidades, Ciencia y Tecnología) program for their Ph.D. studies. This work was supported by CONAHCYT research grants 280789 and 320987. Furthermore, this research was made possible thanks to the generous assistance provided by the Max Planck Institute for Radio Astronomy (MPIfR) - Mexico Max Planck Partner Group led by V.M.P.-A. Data from the Steward Observatory spectropolarimetric monitoring project were used. This program is supported by Fermi Guest Investigator grants NNX08AW56G, NNX09AU10G, NNX12AO93G, and NNX15AU81G. This publication is based on data collected at the Observatorio Astrofísico Guillermo Haro (OAGH), Cananea, Sonora, Mexico, operated by the Instituto Nacional de Astrofísica, Óptica y Electrónica (INAOE). Funding for the OAGH has been provided by CONAHCYT.

\software{Astropy \citep{Astropy2013, Astropy2018, Astropy2022},
	  IRAF \citep{Tody1986, Tody1993},
          SciPy \citep{Virtanen2020}.
          }
          
\hypertarget{paperI}{}
\bibliography{ManuscriptPKS_II}{}
\bibliographystyle{aasjournal}



\end{document}